\documentclass[iop]{emulateapj}

\usepackage{natbib,graphicx,amsmath}

\usepackage[usenames]{color}

\shorttitle{POST-NEWTONIAN MHD}
\shortauthors{HWANG \& NOH}

\newcommand{\bea}{\begin{eqnarray}}
\newcommand{\eea}{\end{eqnarray}}

\begin{document}

\title{Post-Newtonian Magnetohydrodynamics}
\author{Jai-chan Hwang${}^{1,2}$, Hyerim Noh${}^{2,3}$}
\address{${}^{1}$Department of Astronomy and Atmospheric Sciences,
         Kyungpook National University, Daegu, Korea \\
         ${}^{2}$Centre for Theoretical Cosmology, DAMTP, University of Cambridge, CB3 0WA Cambridge, United Kingdom \\
         ${}^{3}$Center for Large Telescope,
         Korea Astronomy and Space Science Institute, Daejon, Korea
         }


\begin{abstract}

Using the fully nonlinear and exact perturbation formulation with magnetohydrodynamics (MHD) in Minkowski background we derive first-order post-Newtonian (1PN) equations without imposing the slicing (temporal gauge) condition. The 1PN MHD formulation is complementary to our recently presented fully relativistic MHD combined with 0PN gravity available only in the maximal slicing. We present the 1PN MHD equations in two gauge conditions previously used in the literature and provide gauge transformation relations between different gauges. We derive the PN effects on MHD waves in a static homogeneous medium.

\end{abstract}

\noindent

\maketitle

\section{Introduction}

Post-Newtonian (PN) approximation is a way of managing the relativistic effects of matter and gravity in the situation where the relativistic effects are weak, thus weak gravity and slow motion. In the PN approximation the relativistic effects are systematically treated as corrections to the well-known non-relativistic (Newtonian) limit.
The PN approximation is based on the expansion in dimensionless parameters like $v^2/c^2$, $GM/(\ell c^2)$ and $\Phi/c^2$ with $v$, $M$, $\ell$ and $\Phi$ the characteristic velocity, mass, length scale, and the gravitational potential, respectively, involved in the system. The expansion on these parameters involving $c^{-2n}$-order is regarded as the $n$th-order PN ($n$PN) expansion (Poisson \& Will 2014).

PN formulation of hydrodynamics was studied by Chandrasekhar and collaborators in a series of papers reaching up to 2.5PN order (Chandrasekhar 1965; Chandrasekhar \& Nutku 1969; Chandrasekhar \& Esposito 1970). 1PN formulation of magnetohydrodynamics (MHD) was presented by Greenberg (1971) and only recently another paper on the subject was presented by Nazari \& Roshan (2018).

The fully nonlinear and exact perturbation (FNLE) formulation of Einstein's gravity with MHD in the Minkowski background is presented recently in Noh, Hwang \& Bucher (2019, NHB hereafter). The formulation is designed to produce nonlinear (higher order) perturbation equations in any temporal gauge (slicing, hypersurface) condition with easy. As the formulation is valid to fully nonlinear order and exact, it may have diverse applications. In NHB we showed that the fully relativistic (special relativistic) MHD can be combined with weak gravity consistently in a certain gauge condition.

In this work we will derive the 1PN approximation of MHD as a complementary formulation. While the fully relativistic MHD combined with weak gravity is possible only in a certain temporal gauge condition (the maximal slicing), our 1PN equations will be presented without fixing the slicing condition. In Hwang, Noh \& Puetzfeld (2008, HNP hereafter) we presented 1PN hydrodynamics in cosmological context without fixing the temporal gauge. Here, we are extending the formulation to include the MHD, but in the Minkowski background.

Section \ref{sec:FN} briefly introduces notations used in the FNLE formulation with MHD, and using the formulation a complete set of equations valid to 1PN order is derived in the Appendix. Section \ref{sec:PN-MHD} presents the 1PN order MHD equations without fixing the temporal gauge condition; in a conventional notation, see Section \ref{sec:Conventional}. Section \ref{sec:Comparison} provides the 1PN equations in two gauge conditions previously used in the literature, and shows the gauge transformation properties between different gauge conditions. Section \ref{sec:waves} presents the PN corrections to the MHD waves in a static homogeneous medium without gravity. Section \ref{sec:Discussion} is a discussion.

We adopt the cgs unit.

\begin{widetext}
\section{Fully nonlinear and exact perturbations with MHD}
                                        \label{sec:FN}

Here we introduce our notations. Details on the FNLE formulation with MHD can be found in NHB. Our metric convention is
\bea
   & & \widetilde g_{00} = - \left( 1 + 2 \alpha \right), \quad
       \widetilde g_{0i} = - \chi_i , \quad
       \widetilde g_{ij} = \left( 1 + 2 \varphi \right) \delta_{ij},
   \label{metric}
\eea
where $\alpha$, $\varphi$ and $\chi_i$ are functions of spacetime with arbitrary amplitudes; the tildes indicate the covariant quantities. The spatial index of $\chi_i$ is raised and lowered using $\delta_{ij}$ as the metric. In this metric convention the spatial part of metric looks simple because we have {\it ignored} the transverse-tracefree (gravitational waves) part of the spatial metric (which is a serious assumption on two physical degrees of freedom), and (without losing generality) imposed a spatial gauge condition (fixing three coordinate degrees of freedom) which removes the spatial gauge mode completely to all perturbation orders; this statement is true as long as we simultaneously choose a temporal gauge condition which removes the temporal gauge mode completely; under these spatial and temporal gauge conditions all remaining variables can be equivalently regarded as (spatially and temporally) gauge invariant ones to all perturbation orders (Bardeen 1988; Section VI of Noh \& Hwang 2004; Hwang \& Noh 2013).

The energy momentum tensor is introduced as
\bea
   & & \widetilde T_{ab}
       = \widetilde \mu \widetilde u_a \widetilde u_a
       + \widetilde p \left( \widetilde g_{ab}
       + \widetilde u_a \widetilde u_b \right)
       + \widetilde \pi_{ab},
\eea
where $\widetilde \mu$, $\widetilde p$ and $\widetilde \pi_{ab}$ are the energy density, pressure and the anisotropic stress ($\widetilde \pi_{ab} \widetilde u^b \equiv 0 \equiv \widetilde \pi^c_c$, $\widetilde \pi_{ab} = \widetilde \pi_{ba}$), respectively, based on the normalized time-like ($\widetilde u^a \widetilde u_a \equiv -1$) four-vector $\widetilde u_a$ in the energy frame [setting the flux term to vanish as $\widetilde q_a \equiv - \widetilde T_{cd} \widetilde u^c ( \widetilde \delta^d_a + \widetilde u^d \widetilde u_a) \equiv 0$]. We introduce the fluid velocity $v_i$ as
\bea
   & & \widetilde u_i \equiv \gamma {v_i \over c}, \quad
       \gamma \equiv {1 \over \sqrt{ 1 - {1 \over 1 + 2 \varphi}
       {v^2 \over c^2}}}, \quad
       v^2 \equiv v^i v_i.
\eea
The spatial index of $v_i$ is raised and lowered using $\delta_{ij}$ as the metric. We set
\bea
   & & \widetilde \mu \equiv \mu \equiv \varrho c^2, \quad
       \varrho \equiv \overline \varrho
       \left( 1 + {1 \over c^2} \Pi \right), \quad
       \widetilde p \equiv p, \quad
       \widetilde \pi_{ij} \equiv \Pi_{ij},
\eea
where $\varrho$, $\overline \varrho$, $\overline \varrho \Pi$ are the density, mass density and the internal energy density, respectively. The fluid quantities are functions of spacetime with arbitrary amplitudes; spatial indices of $\Pi_{ij}$ are raised and lowered using $\delta_{ij}$ as the metric. The complete set of FNLE equations is presented in the Appendix of Hwang \& Noh (2016) for a hydrodynamic fluid.

In the presence of electromagnetism, the energy-momentum tensor of electromagnetic field is
\bea
   & & \widetilde T^{EM}_{ab}
       = {1 \over 4 \pi} \left( \widetilde F_{ac}
       \widetilde F_b^{\;\;c}
       - {1 \over 4} \widetilde g_{ab}
       \widetilde F_{cd} \widetilde F^{cd} \right).
   \label{Tab-EM}
\eea
The electromagnetic tensor can be decomposed as
\bea
   & & \widetilde F_{ab}
       \equiv \widetilde U_a \widetilde E_b
       - \widetilde U_b \widetilde E_a
       - \widetilde \eta_{abcd} \widetilde U^c \widetilde B^d,
   \label{Fab}
\eea
with $\widetilde E_a \widetilde U^a \equiv 0 \equiv \widetilde B_a \widetilde U^a$; $\widetilde U_a$ is a generic normalized time-like four-vector with $\widetilde U^c \widetilde U_c \equiv -1$; it can be the fluid four-vector $\widetilde u_a$ (comoving frame) or the normal four-vector $\widetilde n_a$ (laboratory frame); for the normal four-vector, we have $\widetilde n_i \equiv 0$. For fields in the laboratory frame we introduce
\bea
   & & \widetilde E^{(n)}_i \equiv E_i \equiv {\bf E}, \quad
       \widetilde B^{(n)}_i \equiv B_i \equiv {\bf B},
   \label{E-B}
\eea
where indices of $E_i$ and $B_i$ are raised and lowered using $\delta_{ij}$ as the metric.

The Ohm's law is expressed in the comoving frame as
\bea
   & & \widetilde j_a^{(u)} = \sigma \widetilde E_a^{(u)},
\eea
with $\sigma$ being the electric conductivity. Ideal MHD takes a perfectly conducting limit, $\sigma \rightarrow \infty$ with $\widetilde E_a^{(u)} = 0$, and $\widetilde j_a^{(u)}$ is non-vanishing. From $\widetilde E_a^{(u)} = 0$ we have the ideal MHD condition [see Equation (57) in NHB]
\bea
   & & {\bf E}
        = - {1 \over \sqrt{1 + 2 \varphi}} {1 \over c}
        {\bf v} \times {\bf B}.
   \label{MHD-condition}
\eea
The complete set of FNLE equations with MHD is presented in NHB.

\section{Post-Newtonian approximation with MHD}
                                     \label{sec:PN-MHD}

To 1PN order we set
\bea
   & & \alpha \equiv {\Phi \over c^2}, \quad
       \varphi \equiv - {\Psi \over c^2}, \quad
       \chi_i \equiv {P_i \over c^3}.
   \label{metric-PN}
\eea
Compared with notations used in Chandrasekhar (1965) and Chandrasekhar \& Nutku (1969), we have
\bea
   & & \alpha \equiv {\Phi \over c^2}
       \equiv - {1 \over c^2} \left[
       U + {1 \over c^2} \left( 2 \Upsilon - U^2 \right) \right], \quad
       \varphi \equiv - {\Psi \over c^2}
       \equiv {1 \over c^2} V, \quad
       v_i = \overline v_i
       + {1 \over c^2} \left[
       \left( U + 2 V \right) \overline v_i
       - P_i \right].
   \label{metric-PN-Chandrasekhar}
\eea
Using FNLE notation we have
\bea
   & & {\overline v^i \over c}
       \equiv {\widetilde u^i \over \widetilde u^0}
       = {d x^i \over d x^0},
\eea
where the index of $\overline v_i$ is raised and lowered using $\delta_{ij}$ as the metric. Thus, we have [see the Appendix D in Hwang \& Noh (2013)]
\bea
   & & \widetilde u^i \equiv {\gamma \over N}
       {\overline v^i \over c}, \quad
       \widetilde u^0 = {\gamma \over N}, \quad
       v_i = {1 + 2 \varphi \over N} \overline v_i
       - {c \over N} \chi_i, \quad
       \gamma \equiv {1 \over
       \sqrt{1 - {1 \over 1 + 2 \varphi} {v^k v_k \over c^2}}}, \quad
       N = \sqrt{1 + 2 \alpha
       + {\chi^k \chi_k \over 1 + 2 \varphi}},
   \label{u^i}
\eea
where $\gamma$ is the Lorentz factor and $N$ is the lapse function.

The complete set of Einstein equations, conservation equations and Maxwell equations are {\it derived} in the Appendix by reducing the FNLE equations presented in NHB to 1PN order. Equation (\ref{eq5-PN-MHD}) gives
\bea
   & & \Psi = \Phi,
   \label{eq5-PN}
\eea
to the 0PN order, thus $V = U$. Equation (\ref{eq1-PN-MHD}) gives
\bea
   & & \kappa
       = {1 \over c^2} \left( 3 \dot \Phi
       - P^k_{\;\;,k} \right).
   \label{eq1-PN}
\eea
Using these, Einstein equations in (\ref{eq3-PN-MHD}) and (\ref{eq4-PN-MHD}) give
\bea
   & & \Delta P_i
       - P^k_{\;\;,ki}
       + 4 \dot \Phi_{,i}
       = - 16 \pi G \overline \varrho v_i,
   \label{eq3-PN} \\
   & & \Delta \Phi
       = 4 \pi G \overline \varrho
       + {1 \over c^2} \left[
       4 \pi G \left( \overline \varrho \Pi
       + 2 \overline \varrho v^2 + 3 p \right)
       - 3 \ddot \Phi
       + 2 \Phi^{,k} \Phi_{,k}
       + \dot P^k_{\;\;,k}
       + G B^2 \right].
   \label{eq4-PN}
\eea
The energy, momentum and mass conservation equations in (\ref{eq6-PN-MHD}), (\ref{eq7-PN-MHD}) and (\ref{eq0-PN-MHD}), respectively, give
\bea
   & & {\partial \over \partial t} \left\{ \overline \varrho
       + {1 \over c^2} \left[ \overline \varrho \left( \Pi
       + v^2 - 3 \Phi \right)
       + {1 \over 8 \pi} B^2 \right]
       \right\}
   \nonumber \\
   & & \qquad
       + \nabla^i \left\{
       \overline \varrho v_i
       + {1 \over c^2} \left[
       \overline \varrho \left( \Pi + v^2 \right) v_i
       + p v_i + \Pi_{ij} v^j
       + \overline \varrho P_i
       - {1 \over 4 \pi}
       \left[ \left( {\bf v} \times {\bf B} \right)
       \times {\bf B} \right]_i \right]
       \right\}
       = - {1 \over c^2}
       \overline \varrho {\bf v}
       \cdot \nabla \Phi,
   \label{eq6-PN} \\
   & & {\partial \over \partial t} \left\{
       \overline \varrho v_i
       + {1 \over c^2} \left[
       \overline \varrho \left( \Pi + v^2 - 3 \Phi \right)v_i
       + p v_i
       + \Pi_{ij} v^j
        -{1 \over 4 \pi} \left[ \left( {\bf v} \times {\bf B} \right) \times {\bf B} \right]_i
        \right]
       \right\}
   \nonumber \\
   & & \qquad
       + \nabla^j \bigg\{
       \overline \varrho v_i v_j
       + p \delta_{ij} + \Pi_{ij}
       + {1 \over c^2} \left[ \overline \varrho
       \left( \Pi + v^2 \right) v_i v_j
       + p v_i v_j
       - 2 p \delta_{ij} \Phi
       + \overline \varrho v_i P_j \right]
   \nonumber \\
   & & \qquad
       - {1 \over 4 \pi} \left(
       B_i B_j - {1 \over 2} \delta_{ij} B^2
       \right)
       - {1 \over 4 \pi c^2}
       \left[ \left( {\bf v} \times {\bf B} \right)_i
       \left( {\bf v} \times {\bf B} \right)_j
       - {1 \over 2} \delta_{ij}
       \left| {\bf v} \times {\bf B} \right|^2 \right]
       \bigg\}
   \nonumber \\
   & & \qquad
       = - \overline \varrho \Phi_{,i}
       - {1 \over c^2} \left\{ \Phi_{,i} \left[
       \overline \varrho \left( \Pi + 2 v^2 - 4 \Phi \right)
       + 3 p
       + {B^2 \over 4 \pi} \right]
       + \overline \varrho v_j P^j_{\;\;,i}
       \right\},
   \label{eq7-PN} \\
   & & {\partial \over \partial t}
       \left[ \overline \varrho \left(
       1 + {1 \over 2} {v^2 \over c^2}
       - 3 {\Phi \over c^2} \right) \right]
       + \nabla^i \left[ \overline \varrho v_i
       \left( 1 + {1 \over 2} {v^2 \over c^2}
       \right)
       + \overline \varrho {P_i \over c^2} \right] = 0,
   \label{eq0-PN}
\eea
and Maxwell equations in
(\ref{Maxwell-PN-MHD-1})-(\ref{Maxwell-PN-MHD-4}) give
\bea
   & & \nabla \cdot
       \left[ \left( 1 - {\Phi \over c^2} \right) {\bf B} \right]
       = 0,
   \label{Maxwell-PN-1} \\
   & & {\partial \over \partial t}
       \left[ \left( 1 - {\Phi \over c^2} \right)
       {\bf B} \right]
       = \nabla \times
       \left\{ \left[
       \left( 1 + 2 {\Phi \over c^2} \right)
       {\bf v} + {1 \over c^2} {\bf P} \right]
       \times {\bf B} \right\},
   \label{Maxwell-PN-2} \\
   & & \nabla \cdot \left[ \left( 1 - {\Phi \over c^2} \right)
       {\bf E} \right]
       = - {1 \over c} \nabla \cdot
       \left( {\bf v} \times {\bf B} \right)
       = 4 \pi \varrho_{\rm em} \left( 1 - 3 {\Phi \over c^2} \right),
   \label{Maxwell-PN-3} \\
   & & {\partial \over \partial t}
       \left[ \left( 1 - {\Phi \over c^2} \right)
       {\bf E} \right]
       = - {1 \over c} {\partial \over \partial t}
       \left( {\bf v} \times {\bf B} \right)
       = c \nabla \times
       \left[ \left( 1 + {\Phi \over c^2} \right)
       {\bf B} \right]
       - 4 \pi \left( {\bf j}
       + {1 \over c^2} \varrho_{\rm em} {\bf P} \right).
   \label{Maxwell-PN-4}
\eea
To 1PN order Equation (\ref{MHD-condition}) gives
\bea
   & & {\bf E}
        = - \left( 1 + {\Phi \over c^2} \right)
        {1 \over c} {\bf v} \times {\bf B}.
   \label{MHD-condition-PN}
\eea
Using Maxwell's equations the MHD contributions in Equations (\ref{eq6-PN}) and (\ref{eq7-PN}) can be written on the right-hand-sides, respectively, as
\bea
   & & - {1 \over 4 \pi c^2}
       \left( {\bf v} \times {\bf B} \right) \cdot
       \left( \nabla \times {\bf B} \right),
   \nonumber \\
   & & + {1 \over 4 \pi} \left[
       \left( \nabla \times {\bf B} \right)
       \times {\bf B} \right]_{i}
       + {1 \over 4 \pi c^2}
       \left\{ \left[ \left( {\bf v} \times
       {\bf B} \right)^{\displaystyle{\cdot}} \times {\bf B}
       \right]_i
       + \left( {\bf v} \times {\bf B} \right)_i
       \nabla \cdot \left( {\bf v} \times {\bf B} \right)
       + \left[ {\bf B} \times \left( {\bf B} \times
       \nabla \Phi \right) \right]_i \right\}.
   \label{Ej-PN}
\eea

To 0PN order, the conservation equations in (\ref{eq6-PN})-(\ref{eq0-PN}) give
\bea
   & & \dot {\overline \varrho}
       + \nabla \cdot \left( \overline \varrho {\bf v} \right)
       = 0,
   \label{eq6-0PN} \\
   & & \left( \overline \varrho v_i \right)^{\displaystyle{\cdot}}
       + \nabla^j \left(
       \overline \varrho v_i v_j
       + p \delta_{ij} + \Pi_{ij} \right)
       + \overline \varrho \nabla_i \Phi
       = {1 \over 4 \pi} \left[
       \left( \nabla \times {\bf B} \right)
       \times {\bf B} \right]_{i},
   \label{eq7-0PN}
\eea
thus
\bea
   & & \dot v_i
       + {\bf v} \cdot \nabla v_i
       + {1 \over \overline \varrho}
       \nabla^j \left( p \delta_{ij} + \Pi_{ij} \right)
       + \nabla_i \Phi
       = {1 \over 4 \pi \overline \varrho} \left[
       \left( \nabla \times {\bf B} \right)
       \times {\bf B} \right]_{i}.
   \label{eq7-0PN-2}
\eea

Now, to 1PN order, from Equations (\ref{eq6-PN}) and (\ref{eq0-PN}) we have
\bea
   & & \overline \varrho \left( \dot \Pi
       + {\bf v} \cdot \nabla \Pi \right)
       + p \nabla \cdot {\bf v}
       + \Pi^{ij} v_{i,j} = 0.
   \label{eq6-PN-2}
\eea
Thus, the gravity and MHD do not appear in the internal energy density conservation equation to 1PN order. From Equations (\ref{eq6-PN}) and (\ref{eq7-PN}) we can derive
\bea
   & & \left\{ \overline \varrho
       + {1 \over c^2} \left[ \overline \varrho
       \left( \Pi + v^2 - 3 \Phi \right) + p \right] \right\}
       \left( \dot v_i + {\bf v} \cdot \nabla v_i \right)
       + \left\{ \overline \varrho
       + {1 \over c^2} \left[ \overline \varrho
       \left( \Pi + 2 v^2 - 4 \Phi \right) + p \right] \right\}
       \Phi_{,i}
       + {1 \over c^2} \overline \varrho {\bf v} \cdot
       \left( 3 \Phi \nabla v_i
       - v_i \nabla \Phi \right)
   \nonumber \\
   & & \qquad
       + {1 \over c^2} \overline \varrho \left( v_j P^j_{\;\;,i}
       + P^j v_{i,j} \right)
       + p_{,i} + \Pi^j_{i,j}
       + {1 \over c^2} \left[
       \dot p v_i
       - 2 p_{,i} \Phi
       + \left( \Pi_{ij} v^j \right)^{\displaystyle{\cdot}}
       - v_i \left( \Pi^j_k v^k \right)_{,j} \right]
       = {1 \over 4 \pi} \left[
       \left( \nabla \times {\bf B} \right)
       \times {\bf B} \right]_{i}
   \nonumber \\
   & & \qquad
       + {1 \over 4 \pi c^2}
       \left\{ \left[ \left( {\bf v} \times
       {\bf B} \right)^{\displaystyle{\cdot}} \times {\bf B}
       \right]_i
       + \left( {\bf v} \times {\bf B} \right)_i
       \nabla \cdot \left( {\bf v} \times {\bf B} \right)
       + v_i \left( {\bf v} \times {\bf B} \right) \cdot
       \left( \nabla \times {\bf B} \right)
       + \left[ {\bf B} \times \left( {\bf B} \times
       \nabla \Phi \right) \right]_i \right\}.
   \label{eq7-PN-2}
\eea

The above 1PN equations are presented without imposing the temporal gauge condition. The gauge transformation properties are studied in Section 6 of HNP. The general gauge conditions can be written as [Equation (210) in HNP]
\bea
   & & P^i_{\;\;,i} - n \dot \Phi = 0,
   \label{gauge-PN}
\eea
with the real numbers $n$ covering
\bea
   & & {\rm Chandrasekhar \; (Standard \; PN) \; gauge,
       \; Maximal \; slicing:}               \hskip .22cm
       n = 3,
   \nonumber \\
   & & {\rm Harmonic \; gauge:}              \hskip 6.08cm
       n = 4,
   \nonumber \\
   & & {\rm Transverse\!-\!shear \; gauge:}  \hskip 4.83cm
       n = 0.
\eea
The standard PN gauge was used by Chandrasekhar (1965) and is the same as the maximal slicing setting the trace of extrinsic curvature equal to zero ($K^i_i = \kappa \equiv 0$). The harmonic gauge condition is used in Section 9 of Weinberg (1972). In the transverse-shear gauge we have $P^i_{\;\;,i} \equiv 0$ where $P_i$ is related to the shear of the normal frame, see Equation (42) in HNP. For various gauge conditions used in the literature in the hydrodynamic PN situations, see
Blanchet, Damour \& Sch$\ddot{\rm a}$fer (1990),
Shibata and Asada (1995),
Asada, Shibata and Futamase (1996),
Asada and Futamase (1997),
Racine and Flanagan (2005), and
Poisson and Will (2014).

\section{In conventional PN notation}
                                      \label{sec:Conventional}

In PN literature $\overline {\bf v}$ and $U$ are used often; to 1PN order from Equation (\ref{metric-PN-Chandrasekhar}) we have
\bea
   & & v_i = \overline v_i
       + {1 \over c^2} \left(
       3 U \overline v_i
       - P_i \right), \quad
       \Phi = - U - {1 \over c^2} \left( 2 \Upsilon - U^2 \right),
\eea
and the general PN gauge conditions in Equation (\ref{gauge-PN}) becomes
\bea
   & & P^i_{\;\;,i} + n \dot U = 0.
   \label{gauge-PN-U}
\eea
Einstein's equations in (\ref{eq3-PN}) and (\ref{eq4-PN}) give
\bea
   & & \Delta P_i
       - \left( P^k_{\;\;,k}
       + 4 \dot U \right)_{,i}
       = - 16 \pi G \overline \varrho v_i,
   \label{eq3-PN-U} \\
   & & \Delta U
       + 4 \pi G \overline \varrho
       = - {1 \over c^2} \left[
       2 \Delta \Upsilon
       + 4 \pi G \overline \varrho \left( \Pi
       + 2 v^2
       + 3 {p \over \overline \varrho }
       + 2 U
       + {B^2 \over 4 \pi \overline \varrho}
       \right)
       + 3 \ddot U
       + \dot P^k_{\;\;,k} \right],
   \label{eq4-PN-U}
\eea
and the conservation equations in (\ref{eq6-PN})-(\ref{eq0-PN}) become
\bea
   & & {\partial \over \partial t} \left\{ \overline \varrho
       + {1 \over c^2} \left[ \overline \varrho \left( \Pi
       + v^2 + 3 U \right)
       + {1 \over 8 \pi} B^2 \right]
       \right\}
   \nonumber \\
   & & \qquad
       + \nabla^i \left\{
       \overline \varrho \overline v_i
       + {1 \over c^2} \left[
       \overline \varrho \left( \Pi + v^2 + 3 U \right) v_i
       + p v_i + \Pi_{ij} v^j
       - {1 \over 4 \pi}
       \left[ \left( {\bf v} \times {\bf B} \right)
       \times {\bf B} \right]_i \right]
       \right\}
       = {1 \over c^2}
       \overline \varrho {\bf v}
       \cdot \nabla U,
   \label{eq6-PN-overline} \\
   & & {\partial \over \partial t} \left\{
       \overline \varrho \overline v_i
       + {1 \over c^2} \left[
       \overline \varrho \left( \Pi + v^2 + 6 U \right) v_i
       + p v_i
       + \Pi_{ij} v^j
       - \overline \varrho P_i
        -{1 \over 4 \pi} \left[ \left( {\bf v} \times {\bf B} \right) \times {\bf B} \right]_i
        \right]
       \right\}
   \nonumber \\
   & & \qquad
       + \nabla^j \bigg\{
       \overline \varrho \overline v_i \overline v_j
       + p \delta_{ij} + \Pi_{ij}
       + {1 \over c^2} \left[ \overline \varrho
       \left( \Pi + v^2 + 6 U \right) v_i v_j
       + p v_i v_j
       + 2 p \delta_{ij} U
       - \overline \varrho v_j P_i \right]
   \nonumber \\
   & & \qquad
       - {1 \over 4 \pi} \left(
       B_i B_j - {1 \over 2} \delta_{ij} B^2
       \right)
       - {1 \over 4 \pi c^2}
       \left[ \left( {\bf v} \times {\bf B} \right)_i
       \left( {\bf v} \times {\bf B} \right)_j
       - {1 \over 2} \delta_{ij}
       \left| {\bf v} \times {\bf B} \right|^2 \right]
       \bigg\}
   \nonumber \\
   & & \qquad
       = \overline \varrho U_{,i}
       + {1 \over c^2} \left\{
       2 \overline \varrho \Upsilon_{,i}
       + U_{,i} \left[
       \overline \varrho \left( \Pi + 2 v^2 + 2 U \right)
       + 3 p
       + {B^2 \over 4 \pi} \right]
       - \overline \varrho v_j P^j_{\;\;,i}
       \right\},
   \label{eq7-PN-overline} \\
   & & {\partial \over \partial t}
       \left[ \overline \varrho \left(
       1 + {1 \over 2} {v^2 \over c^2}
       + 3 {U \over c^2} \right) \right]
       + \nabla^i \left[ \overline \varrho \overline v_i
       \left( 1 + {1 \over 2} {v^2 \over c^2}
       + 3 {U \over c^2} \right) \right] = 0.
   \label{eq0-PN-overline}
\eea
The Maxwell's equations in (\ref{Maxwell-PN-1}) and (\ref{Maxwell-PN-2}) give
\bea
   & & \nabla \cdot
       \left[ \left( 1 + {U \over c^2} \right) {\bf B} \right]
       = 0,
   \label{Maxwell-PN-1-U} \\
   & & {\partial \over \partial t}
       \left[ \left( 1 + {U \over c^2} \right)
       {\bf B} \right]
       = \nabla \times
       \left[
       \left( 1 + {U \over c^2} \right)
       \overline {\bf v}
       \times {\bf B} \right],
   \label{Maxwell-PN-2-overline-U}
\eea
and Equation (\ref{eq7-PN-2}) gives
\bea
   & & \left\{ \overline \varrho
       + {1 \over c^2} \left[ \overline \varrho
       \left( \Pi + v^2 + 6 U \right) + p \right] \right\}
       \left( \dot {\overline v}_i + \overline {\bf v} \cdot \nabla \overline v_i \right)
       + p_{,i} + \Pi^j_{i,j}
       - \left\{ \overline \varrho
       + {1 \over c^2} \left[ \overline \varrho
       \left( \Pi + 2 v^2 + 2 U \right) + p \right] \right\}
       U_{,i}
       - {1 \over c^2} 2 \overline \varrho \Upsilon_{,i}
   \nonumber \\
   & & \qquad
       + {1 \over c^2} \overline \varrho
       \left[ \left( 3 \dot U
       + 4 {\bf v} \cdot \nabla U \right) v_i
       - \dot P_i
       - v^j \left( P_{i,j} - P_{j,i} \right) \right]
       + {1 \over c^2} \left[
       \dot p v_i
       + 2 p_{,i} U
       + \left( \Pi_{ij} v^j \right)^{\displaystyle{\cdot}}
       - v_i \left( \Pi^j_k v^k \right)_{,j} \right]
   \nonumber \\
   & & \qquad
       = {1 \over 4 \pi} \left[
       \left( \nabla \times {\bf B} \right)
       \times {\bf B} \right]_{i}
   \nonumber \\
   & & \qquad
       + {1 \over 4 \pi c^2}
       \left\{ \left[ \left( {\bf v} \times
       {\bf B} \right)^{\displaystyle{\cdot}} \times {\bf B}
       \right]_i
       + \left( {\bf v} \times {\bf B} \right)_i
       \nabla \cdot \left( {\bf v} \times {\bf B} \right)
       + v_i \left( {\bf v} \times {\bf B} \right) \cdot
       \left( \nabla \times {\bf B} \right)
       - \left[ {\bf B} \times \left( {\bf B} \times
       \nabla U \right) \right]_i \right\}.
   \label{eq7-PN-2-overline}
\eea
The gauge conditions directly affect only Einstein equations in (\ref{eq3-PN-U}) and (\ref{eq4-PN-U}).

By defining
\bea
   & & \varrho^* \equiv  \overline \varrho \left[
       1 + {1 \over c^2} \left( {1 \over 2} v^2
       + 3 U \right) \right],
\eea
Equation (\ref{eq0-PN-overline}) gives
\bea
   & & {\partial \over \partial t} \varrho^*
       + \nabla \cdot \left( \varrho^* \overline {\bf v} \right)
       = 0.
   \label{PN-continuity}
\eea
Thus, according to Chandrasekhar (1965) ``in the PN approximation, the mass defined in terms of the density $\varrho^*$ is conserved." In FNLE formulation, from the continuity equation, $( \overline \varrho \widetilde u^c )_{;c} = 0$, using Equation (\ref{u^i}) we have
\bea
   & & \varrho^* \equiv \sqrt{-\widetilde g}
       \widetilde u^0 \overline \varrho
       = \sqrt{h} \gamma \overline \varrho,
\eea
where we used $\sqrt{-\widetilde g} = N \sqrt{h}$; $\widetilde g$ is the determinant of $\widetilde g_{ab}$ and $h$ is the determinant of the ADM intrinsic metric tensor, $h_{ij} \equiv \widetilde g_{ij}$.

Chandrasekhar has similarly proved the PN order conservations of the total linear momentum, the total angular momentum and the total energy of the system and has introduced corresponding momentum and energy in the absence of MHD, see Equations (126), (141) and (165), and Equations (128) and (166) in Chandrasekhar (1965), and Chandrasekhar (1969). Corresponding conservation laws in the presence of MHD were studied in Section VII of Greenberg (1971). Although the mass conservation property shown above is independent of the temporal gauge condition (and independent of the presence of MHD), the momentum, angular momentum and energy conservation properties studied in Chandrasekhar (1965) and Greenberg (1971) are presented in the Chandrasekhar gauge.

\section{Comparison with other studies}
                                      \label{sec:Comparison}

\subsection{Chandrasekhar gauge (Standard PN gauge, Maximal slicing)}

In the Chandrasekhar gauge we have $P^i_{\;\;,i} = - 3 \dot U$. Equations (\ref{eq3-PN-U}) and (\ref{eq4-PN-U}) give
\bea
   & & \Delta P_i
       = - 16 \pi G \overline \varrho v_i
       + \dot U_{,i},
   \label{eq3-PN-CG} \\
   & & \Delta U
       + 4 \pi G \overline \varrho
       = - {1 \over c^2} \left[
       2 \Delta \Upsilon
       + 4 \pi G \overline \varrho \left( \Pi
       + 2 v^2
       + 3 {p \over \overline \varrho }
       + 2 U
       + {B^2 \over 4 \pi \overline \varrho}
       \right)
       \right].
   \label{eq4-PN-CG}
\eea
In the notation of Chandrasekhar (1965), we have
\bea
   & & \Upsilon = \overline \Phi, \quad
       {\bf P} = 4 {\bf U}
       - {1 \over 2} \nabla \dot {\overline \chi},
\eea
thus, metric becomes
\bea
   & & \widetilde g_{00} = - 1 + {2 \over c^2} U
       + {2 \over c^4} \left( 2 \overline \Phi - U^2 \right), \quad
       \widetilde g_{0i} = - {1 \over c^3} \left(
       4 U_i - {1 \over 2} \dot {\overline \chi}_{,i} \right), \quad
       \widetilde g_{ij} = \left( 1 + {2 \over c^2} U \right) \delta_{ij}.
   \label{metric-CG}
\eea
In order to distinguish from our notation we put overlines in Chandrasekhar's $\Phi$ and $\chi$. Chandrasekhar has defined $\overline \chi$ so that we have [Equation (44) in Chandrasekhar (1965)]
\bea
   & & \Delta \overline \chi \equiv - 2 U.
   \label{chi-definition}
\eea
Thus, the Chandrasekhar gauge condition gives $\nabla \cdot {\bf U} = - \dot U$.
Using Equation (\ref{chi-definition}) and the gauge condition, Equation (\ref{eq3-PN-CG}) gives [Equation (45) in Chandrasekhar (1965)]
\bea
   & & \Delta {\bf U} = - 4 \pi G \overline \varrho {\bf v}.
   \label{U_i-Poisson}
\eea
To each PN order, Equation (\ref{eq4-PN-CG}) gives [Equations (3) and (41) in Chandrasekhar (1965)]
\bea
   & & \Delta U
       = - 4 \pi G \overline \varrho,
   \label{U-Poisson}
   \\
   & & \Delta \overline \Phi
       = - 4 \pi G \overline \varrho \left( {1 \over 2} \Pi
       + v^2
       + {3 \over 2} {p \over \overline \varrho }
       + U
       + {B^2 \over 8 \pi \overline \varrho}
       \right)
       \equiv - 4 \pi G \overline \varrho \phi.
   \label{Phi-Poisson}
\eea
The variable $U$ is the Newtonian potential and $\overline \chi$, $U_i$ and $\overline \Phi$ are post-Newtonian potentials introduced in Chandrasekhar (1965), now extended to include the MHD effect (Greenberg 1971). The potentials can be expressed in terms of integrals as [Equations (69) and (82) in Chandrasekhar (1965)]
\bea
   & & U (t, {\bf x})
       = G \int_V {\overline \varrho (t, {\bf x}^\prime)
       \over | {\bf x} - {\bf x}^\prime |} d^3 {\bf x}^\prime, \quad
       {\bf U} (t, {\bf x})
       = G \int_V {\overline \varrho (t, {\bf x}^\prime)
       {\bf v} (t, {\bf x}^\prime)
       \over | {\bf x} - {\bf x}^\prime |} d^3 {\bf x}^\prime, \quad
       \overline \Phi (t, {\bf x})
       = G \int_V {\overline \varrho (t, {\bf x}^\prime)
       \phi (t, {\bf x}^\prime)
       \over | {\bf x} - {\bf x}^\prime |} d^3 {\bf x}^\prime,
   \nonumber \\
   & & \overline \chi (t, {\bf x})
       = - G \int_V \overline \varrho (t, {\bf x}^\prime)
       | {\bf x} - {\bf x}^\prime | d^3 {\bf x}^\prime.
   \label{Potential-solutions}
\eea
The gauge condition does not directly affect the conservation equations and Maxwell's equations in (\ref{eq6-PN-overline})-(\ref{eq7-PN-2-overline}).

\subsection{Harmonic gauge}

In the harmonic gauge we have $P^i_{\;\;,i} = - 4 \dot U$. Equations (\ref{eq3-PN-U}) and (\ref{eq4-PN-U}) give
\bea
   & & \Delta P_i
       = - 16 \pi G \overline \varrho v_i,
   \label{eq3-PN-HG} \\
   & & \Box U
       + 4 \pi G \overline \varrho
       = - {1 \over c^2} \left[
       2 \Delta \Upsilon
       + 4 \pi G \overline \varrho \left( \Pi
       + 2 v^2
       + 3 {p \over \overline \varrho }
       + 2 U
       + {B^2 \over 4 \pi \overline \varrho}
       \right)
       \right].
   \label{eq4-PN-HG}
\eea
In this gauge the propagation speed of the potential $U$ is the speed of light, whereas in the Chandrasekhar gauge all potentials satisfy Poisson-like equations as in (\ref{chi-definition})-(\ref{Phi-Poisson}) with action-at-a-distance nature. Using $\overline \chi$ defined in Equation (\ref{chi-definition}), and by introducing
\bea
   & & \Upsilon = \overline \Phi
       - {1 \over 4} \ddot {\overline \chi}, \quad
       {\bf P} = 4 {\bf U},
\eea
Equations (\ref{eq3-PN-HG}) and (\ref{eq4-PN-HG}) give exactly the same equations in (\ref{U_i-Poisson})-(\ref{Phi-Poisson}), thus solutions in Equation (\ref{Potential-solutions}) remain valid; we have $\nabla \cdot {\bf U} = - \dot U$ in the harmonic gauge as well. The metric in the harmonic gauge becomes
\bea
   & & \widetilde g_{00} = - 1 + {2 \over c^2} U
       + {2 \over c^4} \left( 2 \overline \Phi
       - {1 \over 2} \ddot {\overline \chi}
       - U^2 \right), \quad
       \widetilde g_{0i} = - {4 \over c^3} U_i , \quad
       \widetilde g_{ij} = \left( 1 + {2 \over c^2} U \right) \delta_{ij}.
   \label{metric-HG}
\eea
The PN MHD formulation in the harmonic gauge was studied in Nazari \& Roshan (2018).

\subsection{Gauge transformation properties}

The gauge transformation properties to 1PN order were studied in Section 6 of HNP. Here we summarize the gauge transformation properties in HNP and expand the case to include MHD. We consider a gauge transformation $\widehat x^a = x^a + \widetilde \xi^a (x^c)$ with $x^0 = c t$ and
\bea
   & & \widetilde \xi^0 \equiv {1 \over c} \xi^{(2)0}
       + {1 \over c^3} \xi^{(4)0} + \dots, \quad
       \widetilde \xi^i \equiv {1 \over c^2} \xi^{(2)i} + \dots,
\eea
where index of $\xi^{(2)i}$ is raised and lowered by $\delta_{ij}$. The spatial gauge condition taken in Equation (\ref{metric}) to simplify the space-space part of the metric leads to $\xi^{(2)0} = 0 = \xi^{(2)i}$, see Equations (171) and (173) in HNP. To 1PN order we have
\bea
   & & \widehat \Upsilon
       = \Upsilon
       + {1 \over 2} \dot \xi^{(4)0}, \quad
       \widehat P_i = P_i
       - \xi^{(4)0}_{\;\;\;\;\;\;,i},
\eea
and the other PN variables are gauge invariant, see Equation (180) in HNP.

The electromagnetic part is a new degree of freedom in addition to hydrodynamic case considered in HNP. To 1PN order, from Equations (\ref{Fab}) and (\ref{E-B}), we have
\bea
   & & \widetilde F_{ij}
       = \left( 1 - {\Phi \over c^2} \right)
       \eta_{ijk} B^k, \quad
       \widetilde F_{0i}
       = - \left( 1 + {\Phi \over c^2} \right) E_i
       + {1 \over c^3} \left( {\bf P} \times {\bf B} \right)_i,
\eea
with $\Phi = - U$ here. Using the tensorial nature of $\widetilde F_{ab}$, and the gauge transformation property of the second-rank tensor presented in Equation (157) of HNP, we can show $\widehat {\widetilde F}_{ij} = \widetilde F_{ij}$ and $\widehat {\widetilde F}_{0i} = \widetilde F_{0i}$ to 1PN order. Thus we have
\bea
   & & \widehat {\bf B} = {\bf B}, \quad
       \widehat {\bf E} = {\bf E} + {1 \over c^3}
       {\bf B} \times \nabla \xi^{(4)0}.
\eea

\subsection{Gauge transformation between Chandrasekhar gauge and harmonic gauge}

Using the gauge transformation properties, we can relate 1PN variables between the Chandrasekhar gauge (CG) and the harmonic gauge (HG). In $\widehat t_{\rm CG} = t_{\rm HG} + {1 \over c^4} \xi^{(4)0}_{\rm HG \rightarrow CG}$ we consider the hat coordinate to be the Chandrasekhar gauge and the non-hat coordinate to be the harmonic gauge. From
\bea
   & & \widehat P_i |_{\rm CG}
       = 4 U_i - {1 \over 2} \dot {\overline \chi}_{,i}, \quad
       P_i |_{\rm HG} = 4 U_i, \quad
       \widehat \Upsilon |_{\rm CG} = \overline \Phi, \quad
       \Upsilon |_{\rm HG} = \overline \Phi - {1 \over 4} \ddot {\overline \chi},
\eea
we have $\xi^{(4)0}_{\rm HG \rightarrow CG} = {1 \over 2} \dot {\overline \chi}$. Thus we have
\bea
   & & \widehat P_i |_{\rm CG}
       = P_i |_{\rm HG}
       - {1 \over 2} \dot {\overline \chi}_{,i}, \quad
       \widehat \Upsilon |_{\rm CG}
       = \Upsilon |_{\rm HG}
       + {1 \over 4} \ddot {\overline \chi}, \quad
       \widehat {\bf E} |_{\rm CG}
       = {\bf E} |_{\rm HG}
       + {1 \over 2 c^3} {\bf B} \times \nabla \dot {\overline \chi},
\eea
and other 1PN variables are gauge invariant.

\section{PN MHD waves}
                                         \label{sec:waves}

Here we present 1PN corrections to the MHD waves. We follow Section 22 of Shu (1992) which present the case without PN correction.

We {\it ignore} the internal energy, stress and the gravity (thus $\overline {\bf v} = {\bf v}$), and {\it consider} a static and homogeneous background medium with
\bea
   & & \overline \varrho_0 = {\rm constant}, \quad
       p_0 = {\rm constant}, \quad
       {\bf B}_0 = B_0 \widehat {\bf n} = {\rm constant}, \quad
       {\bf v}_0 = 0,
\eea
where $\widehat {\bf n}$ is a constant unit vector. Introducing perturbations as
\bea
   & & \overline \varrho
       = \overline \varrho_0 + \delta \overline \varrho
       = \overline \varrho_0 \left( 1 + \overline \delta \right), \quad
       p = p_0 + \delta p, \quad
       {\bf B} = {\bf B}_0 + \delta {\bf B},
\eea
to the linear order perturbation, Equations (\ref{eq0-PN-overline})-(\ref{eq7-PN-2-overline}) give
\bea
   & & \dot {\overline \delta} + \nabla \cdot {\bf v} = 0, \quad
   \label{continuity-Alfven-pert} \\
   & & \left( \overline \varrho_0
       + {p_0 \over c^2} \right) \dot {\bf v}
       + \nabla \delta p
       = {1 \over 4 \pi} \left[
       \left( \nabla \times \delta {\bf B} \right) \times {\bf B}_0
       + {1 \over c^2} \left( \dot {\bf v} \times {\bf B}_0
       \right) \times {\bf B}_0
       \right],
   \label{Euler-Alfven-pert} \\
   & & \delta \dot {\bf B}
       = \nabla \times \left( {\bf v} \times {\bf B}_0 \right),
   \label{Maxwell-1-Alfven-pert} \\
   & & \nabla \cdot \delta {\bf B} = 0.
   \label{Maxwell-2-Alfven-pert}
\eea
The PN corrections appear only in the momentum conservation equation in (\ref{Euler-Alfven-pert}); to 1PN order this can be written as
\bea
   & & \left( \overline \varrho_0 + {p_0 \over c^2}
       + {B_0^2 \over 4 \pi c^2} \right) \dot {\bf v}
       = - \nabla \delta p
       + {B_0 \over 4 \pi}
       \left( \nabla \times \delta {\bf B} \right)
       \times \widehat {\bf n}
       - {B_0^2 \over 4 \pi \overline \varrho_0 c^2}
       \widehat {\bf n} \widehat {\bf n} \cdot \nabla \delta p.
   \label{Euler-Alfven-pert-1}
\eea
We consider perturbation variables depending on Fourier expansion $e^{i ({\bf k} \cdot {\bf x} - \omega t)}$, thus
\bea
   & & i \omega \overline \delta = i {\bf k} \cdot {\bf v},
   \label{continuity-Alfven-pert-2} \\
   & & i \omega
       \left( 1 + {p_0 \over \overline \varrho_0 c^2}
       + {c_A^2 \over c^2} \right) {\bf v}
       = i {\bf k} c_s^2 \overline \delta
       - i {B_0 \over 4 \pi \overline \varrho_0}
       \left( {\bf k} \times \delta {\bf B} \right)
       \times \widehat {\bf n}
       + i {c_A^2 c_s^2 \over c^2}
       \widehat {\bf n} \widehat {\bf n} \cdot {\bf k} \delta,
   \label{Euler-Alfven-pert-2} \\
   & & i \omega \delta {\bf B}
       = - i B_0 {\bf k} \times
       \left( {\bf v} \times \widehat {\bf n} \right),
   \label{Maxwell-1-Alfven-pert-2} \\
   & & i {\bf k} \cdot \delta {\bf B} = 0,
   \label{Maxwell-2-Alfven-pert-2}
\eea
where we introduced the adiabatic sound velocity $c_s$ and the Alfven velocity $c_A$ as
\bea
   & & c_s^2 \equiv {\delta p \over \delta \overline \varrho}, \quad
       c_A^2 \equiv {B_0^2 \over 4 \pi \overline \varrho_0}.
\eea
Equation (\ref{Maxwell-1-Alfven-pert-2}) implies Equation (\ref{Maxwell-2-Alfven-pert-2}). Combining Equations (\ref{continuity-Alfven-pert-2})-(\ref{Maxwell-1-Alfven-pert-2}), we have
\bea
   & & \left[ \omega^2
       \left( 1 + {p_0 \over \overline \varrho_0 c^2}
       + {c_A^2 \over c^2} \right)
       - c_A^2 \left( {\bf k} \cdot \widehat {\bf n} \right)^2
       \right] {\bf v}
       = {\bf k} \left[ \left( c_s^2 + c_A^2 \right)
       {\bf k} \cdot {\bf v}
       - c_A^2 {\bf v} \cdot \widehat {\bf n}
       {\bf k} \cdot \widehat {\bf n} \right]
       - \widehat {\bf n} c_A^2 {\bf k} \cdot \widehat {\bf n}
       {\bf k} \cdot {\bf v}
       \left( 1 - {c_s^2 \over c^2} \right).
\eea
Following Shu (1992) we set the coordinate as
\bea
   & & {\bf k} \equiv k \widehat {\bf x}, \quad
       \widehat {\bf n} \equiv \cos{\psi} \widehat {\bf x}
       + \sin{\psi} \widehat {\bf y}, \quad
       {\bf v} \equiv v_x \widehat {\bf x}
       + v_y \widehat {\bf y} + v_z \widehat {\bf z}.
\eea
Thus, we have
\bea
   & & \left[ \omega^2
       \left( 1 + {p_0 \over \overline \varrho_0 c^2}
       + {c_A^2 \over c^2} \right)
       - k^2 c_A^2 \cos^2{\psi}
       \right] v_z = 0,
   \label{v_z} \\
   & &
       \left(
       \begin{array}{cc}
           \omega^2 \left( 1 + {p_0 \over \overline \varrho_0 c^2}
           + {c_A^2 \over c^2} \right) - k^2 \left( c_s^2
           + c_A^2 \sin^2{\psi}
           + {c_A^2 c_s^2 \over c^2} \cos^2{\psi} \right)
           & k^2 c_A^2 \sin{\psi} \cos{\psi}
       \\
           k^2 c_A^2 \sin{\psi} \cos{\psi}
           \left( 1 - {c_s^2 \over c^2} \right)
           & \omega^2 \left( 1 + {p_0 \over \overline \varrho_0 c^2}
           + {c_A^2 \over c^2} \right)
           - k^2 c_A^2 \cos^2{\psi}
       \end{array}
       \right)
       \left(
       \begin{array}{c}
           v_x
       \\
           v_y
       \end{array}
       \right)
       =
       \left(
       \begin{array}{c}
           0
       \\
           0
       \end{array}
       \right).
   \label{v_xy}
\eea

For $v_z \neq 0$ (${\bf v}$ perpendicular to ${\bf B_0}$-${\bf k}$ plane), from Equation (\ref{v_z}) we have the Alfven waves with the velocity
\bea
   & & {\omega^2 \over k^2}
       = c_A^2 \cos^2{\psi} \left( 1 - {p_0 \over \overline \varrho_0 c^2}
       - {c_A^2 \over c^2} \right).
   \label{Alfven-wave}
\eea
Notice the PN corrections reduce the velocity.

For non-vanishing $v_x$ and $v_y$ (${\bf v}$ in ${\bf B_0}$-${\bf k}$ plane), from Equation (\ref{v_xy}) we have
\bea
   & & \left[ {\omega^2 \over k^2}
       \left( 1 + {p_0 \over \overline \varrho_0 c^2}
       + {c_A^2 \over c^2} \right) \right]^2
       - \left[ {\omega^2 \over k^2}
       \left( 1 + {p_0 \over \overline \varrho_0 c^2}
       + {c_A^2 \over c^2} \right) \right]
       \left( c_s^2 + c_A^2
       + {c_A^2 c_s^2 \over c^2} \cos^2{\psi} \right)
       + c_A^2 c_s^2 \cos^2{\psi}
       \left( 1 + {c_A^2 \over c^2} \right) = 0,
\eea
with solutions
\bea
   & & {\omega^2 \over k^2}
       \left( 1 + {p_0 \over \overline \varrho_0 c^2}
       + {c_A^2 \over c^2} \right)
       = {1 \over 2} \left[ c_s^2 + c_A^2
       + {c_A^2 c_s^2 \over c^2} \cos^2{\psi}
       \pm \sqrt{ \left( c_s^2 + c_A^2
       + {c_A^2 c_s^2 \over c^2} \cos^2{\psi} \right)^2
       - 4 c_A^2 c_s^2 \cos^2{\psi}
       \left( 1 + {c_A^2 \over c^2} \right)} \right].
\eea
The plus and minus signs in $\pm$ correspond to the fast and slow waves, respectively.
For ${\bf k} \parallel {\bf B}_0$ ($\psi = 0^{\rm o}$), we have
\bea
   & & {\omega^2 \over k^2}
       = c_s^2 \left( 1 - {p_0 \over \overline \varrho_0 c^2} \right), \quad
       c_A^2 \left( 1 - {p_0 \over \overline \varrho_0 c^2}
       - {c_A^2 \over c^2} \right),
   \label{fast-slow-wave}
\eea
with the faster (slower) mode the fast (slow) MHD waves (Shu 1992). For ${\bf k} \perp {\bf B}_0$ ($\psi = 90^{\rm o}$), we have
\bea
   & & {\omega^2 \over k^2}
       = \left( c_s^2 + c_A^2 \right)
       \left( 1 - {p_0 \over \overline \varrho_0 c^2}
       - {c_A^2 \over c^2} \right), \quad
       0,
   \label{magnetosonic-wave}
\eea
with the fast mode the magnetosonic wave (Shu 1992) and the slow mode vanishing. Equations (\ref{Alfven-wave}), (\ref{fast-slow-wave}) and (\ref{magnetosonic-wave}) show that the PN effects of the pressure and the magnetic pressure, $p_{\rm M} \equiv B^2/(8 \pi)$, of the background tend to slowdown all the wave propagation velocities. The above analysis is gauge invariant.

\section{Discussion}
                                         \label{sec:Discussion}

We presented general relativistic MHD equations valid to 1PN order, (Sections \ref{sec:PN-MHD}-\ref{sec:Comparison}). Derivation is presented in the Appendix using the FNLE formulation with MHD shown in NHB. Our 1PN-MHD formulation is complementary to the special relativistic (SR) MHD combined with the weak gravity presented also in NHB. Our 1PN-MHD considers 1PN order expansion for matter, field and gravity consistently. Whereas, the SR-MHD with weak gravity considers fully relativistic (thus $\infty$PN) order in matter and field matched with non-relativistic (thus 0PN) order in gravity. It is not {\it a priori} obvious that such an asymmetric combination is possible. In NHB we have shown that all equations in Einstein's gravity are consistently valid with such a combination in the maximal slicing, see Hwang \& Noh (2016) in the hydrodynamic situation. Our 1PN-MHD formulation is presented without imposing the temporal gauge condition; for general gauge conditions see Equation (\ref{gauge-PN}). 1PN approximation including the ideal MHD in Minkowski background is studied by Greenberg (1971) in the Chandrasekhar gauge and by Nazari \& Roshan (2018) in the harmonic gauge. Comparisons are made in Section \ref{sec:Comparison}. The PN corrections to the well-known MHD waves in a static homogeneous medium without gravity are presented in Section \ref{sec:waves}; to 1PN order the gas pressure as well as the magnetic pressure tend to slow down the wave speeds.

Considering the fully nonlinear and exact nature of the original formulation in NHB, it is a trivial procedure to derive higher order PN expansion. The formulation presented in NHB {\it took} a special but unique spatial gauge condition without losing any generality or advantage, but {\it ignored} the transverse-tracefree (TT) perturbation in the spatial metric; ignoring the TT mode is a serious physical restriction excluding the gravitational waves. But these two assumptions were completely relaxed in Gong et al (2017) in the cosmological context; by setting the scale factor to be unity and ignoring the cosmological constant, we recover the formulation in Minkowski background. Thus, our FNLE formulation may provide easier route to derive higher order PN expansion as well as higher order perturbation equations.

The geodesic equations for dust particles (time-like) and photons (null-like) are presented in Section 5 of HNP in the context of cosmology. The presence of MHD does not affect the geodesic equations.

%
%
\section*{Acknowledgments}
H.N.\ was supported by National Research Foundation of Korea funded by the Korean Government (No.\ 2018R1A2B6002466).
J.H.\ was supported by Basic Science Research Program through the National Research Foundation (NRF) of Korea funded by the Ministry of Science, ICT and future Planning (No.\ 2016R1A2B4007964, No.\ 2018R1A6A1A06024970 and NRF-2019R1A2C1003031).

\appendix
\section{Fully nonlinear and exact equations to 1PN order}

Using the 1PN notation in Equation (\ref{metric-PN}), and the ideal MHD condition in Equation (\ref{MHD-condition}), to 1PN order Einstein's and conservation equations in (79)-(86) of NHB give
\bea
   & & \kappa
       = {1 \over c^2} \left( 3 \dot \Psi
       - P^k_{\;\;,k} \right),
   \label{eq1-PN-MHD} \\
   & & - 4 \pi G \overline \varrho
       + \Delta \Psi
       = {\cal O} (c^{-2}),
   \label{eq2-PN-MHD} \\
   & & \kappa_{,i}
       + {3 \over 4 c^2} \left( \Delta P_i
       + {1 \over 3} P^k_{\;\;,ki}
       + 16 \pi G
       \overline \varrho v_i \right)
       = 0,
   \label{eq3-PN-MHD} \\
   & & - 4 \pi G \overline \varrho
       + \Delta \Phi
       = - \dot \kappa
       + {1 \over c^2} \left[
       4 \pi G \left( \overline \varrho \Pi
       + 2 \overline \varrho v^2 + 3 p \right)
       + 2 \left( \Phi - \Psi \right) \Delta \Phi
       + \Phi^{,k} \left( \Phi
       + \Psi \right)_{,k}
       + G B^2 \right],
   \label{eq4-PN-MHD} \\
   & & \left( \nabla^i \nabla_j
       - {1 \over 3} \delta^i_j \Delta \right)
       \left( \Phi - \Psi \right) =  {\cal O} (c^{-2}),
   \label{eq5-PN-MHD} \\
   & & {\partial \over \partial t} \left\{ \overline \varrho
       + {1 \over c^2} \left[ \overline \varrho \left( \Pi
       + v^2 - 3 \Psi \right)
       + {1 \over 8 \pi} B^2 \right]
       \right\}
   \nonumber \\
   & & \qquad
       + \nabla^i \left\{
       \overline \varrho v_i
       + {1 \over c^2} \left[
       \overline \varrho \left( \Pi + v^2
       + \Phi - \Psi \right) v_i
       + p v_i + \Pi_{ij} v^j
       + \overline \varrho P_i
       - {1 \over 4 \pi}
       \left[ \left( {\bf v} \times {\bf B} \right)
       \times {\bf B} \right]_i \right]
       \right\}
       = - {1 \over c^2}
       \overline \varrho {\bf v}
       \cdot \nabla \Phi,
   \label{eq6-PN-MHD} \\
   & & {\partial \over \partial t} \left\{
       \overline \varrho v_i
       + {1 \over c^2} \left[
       \overline \varrho \left( \Pi + v^2 - 3 \Psi \right)v_i
       + p v_i
       + \Pi_{ij} v^j
        -{1 \over 4 \pi} \left[ \left( {\bf v} \times {\bf B} \right) \times {\bf B} \right]_i
        \right]
       \right\}
   \nonumber \\
   & & \qquad
       + \nabla^j \bigg\{
       \overline \varrho v_i v_j
       + p \delta_{ij} + \Pi_{ij}
       + {1 \over c^2} \left[ \overline \varrho
       \left( \Pi + v^2 + \Phi - \Psi \right) v_i v_j
       + p v_i v_j
       + p \delta_{ij} \left( \Phi - 3 \Psi \right)
       + \Pi_{ij} \left( \Phi - \Psi \right)
       + \overline \varrho v_i P_j \right]
   \nonumber \\
   & & \qquad
       - {1 \over 4 \pi} \left(
       B_i B_j - {1 \over 2} \delta_{ij} B^2
       \right)
       \left( 1 + {\Phi \over c^2} - {\Psi \over c^2} \right)
       - {1 \over 4 \pi c^2}
       \left[ \left( {\bf v} \times {\bf B} \right)_i
       \left( {\bf v} \times {\bf B} \right)_j
       - {1 \over 2} \delta_{ij}
       \left| {\bf v} \times {\bf B} \right|^2 \right]
       \bigg\}
   \nonumber \\
   & & \qquad
       = - \overline \varrho \Phi_{,i}
       - {1 \over c^2} \left[ \Phi_{,i}
       \overline \varrho \left( \Pi + v^2 - \Phi - 3 \Psi \right)
       + \Psi_{,i} \left( \overline \varrho v^2 + 3 p \right)
       + \overline \varrho v_j P^j_{\;\;,i}
       + {B^2 \over 8 \pi} \left( \Phi + \Psi \right)_{,i}
       \right],
   \label{eq7-PN-MHD} \\
   & & {\partial \over \partial t}
       \left[ \overline \varrho \left(
       1 + {1 \over 2} {v^2 \over c^2}
       - 3 {\Psi \over c^2} \right) \right]
       + \nabla_i \left[ \overline \varrho v^i
       \left( 1 + {1 \over 2} {v^2 \over c^2}
       + {\Phi \over c^2} - {\Psi \over c^2}
       \right)
       + \overline \varrho {P^i \over c^2} \right] = 0.
   \label{eq0-PN-MHD}
\eea
We used Equation (\ref{eq1-PN-MHD}) in deriving Equation (\ref{eq0-PN-MHD}). We note that terms on the right-hand-sides of Equations (\ref{eq2-PN-MHD}) and (\ref{eq5-PN-MHD}) are 2PN order, thus can be ignored to 1PN order; the FNLE equations presented in NHB, being fully nonlinear, can be expanded to any/all PN orders (except that we {\it ignored} the transverse-tracefree perturbation), and the correct 1PN orders in Equations (\ref{eq2-PN-MHD}) and (\ref{eq5-PN-MHD}) are limited by the spatial curvature terms $R^{(h)}_{ij}$, see Equations (54), (87), (89), (90), (94) and (97) in HNP.

The Maxwell's equations in Equations (87)-(90) of NHB give
\bea
   & & \nabla \cdot
       \left[ \left( 1 - {\Psi \over c^2} \right) {\bf B} \right]
       = 0,
   \label{Maxwell-PN-MHD-1} \\
   & & {\partial \over \partial t}
       \left[ \left( 1 - {\Psi \over c^2} \right)
       {\bf B} \right]
       = \nabla \times
       \left\{ \left[
       \left( 1 + {\Phi \over c^2} + {\Psi \over c^2} \right)
       {\bf v} + {1 \over c^2} {\bf P} \right]
       \times {\bf B} \right\},
   \label{Maxwell-PN-MHD-2} \\
   & & \nabla \cdot \left[ \left( 1 - {\Psi \over c^2} \right)
       {\bf E} \right]
       = - {1 \over c} \nabla \cdot
       \left( {\bf v} \times {\bf B} \right)
       = 4 \pi \varrho_{\rm em} \left( 1 - 3 {\Psi \over c^2} \right),
   \label{Maxwell-PN-MHD-3} \\
   & & {\partial \over \partial t}
       \left[ \left( 1 - {\Psi \over c^2} \right)
       {\bf E} \right]
       = - {1 \over c} {\partial \over \partial t}
       \left( {\bf v} \times {\bf B} \right)
       = c \nabla \times
       \left[ \left( 1 + {\Phi \over c^2} \right)
       {\bf B} \right]
       - 4 \pi \left[ {\bf j}
       \left( 1 + {\Phi \over c^2} - {\Psi \over c^2} \right)
       + {1 \over c^2} \varrho_{\rm em} {\bf P} \right].
   \label{Maxwell-PN-MHD-4}
\eea
Using Maxwell's equations, the MHD contributions in Equations (\ref{eq6-PN-MHD}) and (\ref{eq7-PN-MHD}) can be collected on the right-hand-sides, respectively, as
\bea
   & & - {1 \over 4 \pi c^2}
       \left( {\bf v} \times {\bf B} \right) \cdot
       \left( \nabla \times {\bf B} \right),
   \nonumber \\
   & & + {1 \over 4 \pi} \left[
       \left( \nabla \times {\bf B} \right)
       \times {\bf B} \right]_{i}
       \left( 1 + {\Phi \over c^2} - {\Psi \over c^2} \right)
       + {1 \over 4 \pi c^2}
       \left\{ \left[ \left( {\bf v} \times
       {\bf B} \right)^{\displaystyle{\cdot}} \times {\bf B}
       \right]_i
       + \left( {\bf v} \times {\bf B} \right)_i
       \nabla \cdot \left( {\bf v} \times {\bf B} \right)
       + \left[ {\bf B} \times \left( {\bf B} \times
       \nabla \Phi \right) \right]_i \right\}.
   \label{Ej-PN-App}
\eea
In the absence of the MHD, using the notations in Equation (\ref{metric-PN-Chandrasekhar})
our equations above reproduce 1PN equations in HNP: using Equation (\ref{eq1-PN-MHD}), Equations (\ref{eq3-PN-MHD}), (\ref{eq4-PN-MHD}), (\ref{eq6-PN-MHD}), (\ref{eq7-PN-MHD}) and (\ref{eq0-PN-MHD}), respectively, give Equations (79), (78), (57), (58) and (62) in HNP; by setting the scale factor $a \equiv 1$ and setting the cosmological constant $\Lambda \equiv 0$ in HNP we have the 1PN formulation in Minkowski background.

\end{widetext}
%
%



\begin{thebibliography}{21}
\expandafter\ifx\csname natexlab\endcsname\relax\def\natexlab#1{#1}\fi
\bibitem{Asada-Futamase-1997}
         Asada H., Futamase T., 1997, Prog. Theor. Phys. Suppl., 128, 123
\bibitem{Asada-Shibata-Futamase-1996}
         Asada H., Shibata M., Futamase T., 1996, Prog. Theor. Phys., 96, 81
\bibitem{Blanchet-1990}
         Blanchet L., Damour T. and Sch$\ddot{\rm a}$fer G., 1990 Mon. Not. R. Astron. Soc. 244 289
\bibitem{Chandrasekhar-1965}
         Chandrasekhar S., 1965, ApJ, 142, 1488
\bibitem{Chandrasekhar-1969}
         Chandrasekhar S., 1969, ApJ, 158, 45
\bibitem{Chandrasekhar-Esposito-1970}
         Chandrasekhar S., Esposito F. P., 1970, ApJ, 160, 153
\bibitem{Chandrasekhar-Nutku-1969}
         Chandrasekhar S., Nutku, Y., 1969, ApJ, 158, 55
\bibitem{Gong-2017}
         Gong J., et al., 2017, JCAP, 10, 027
\bibitem{Greenberg-1971}
         Greenberg P. J., 1971, ApJ, 164, 589
\bibitem{Hwang-Noh-2013}
         Hwang J., Noh H. 2013, MNRAS, 433, 3472
\bibitem{SRH-2016}
         Hwang J., Noh H., 2016, ApJ, 833, 180
\bibitem{PN-2008}
         Hwang J., Noh H., Puetzfeld D., 2008, JCAP, 03, 010 (HNP)
\bibitem{Nazari-Roshan-2018}
         Nazari E., Roshan M., 2018, ApJ, 868, 98
\bibitem{Noh-Hwang-2004}
         Noh H., Hwang J., 2004, Phys. Rev. D, 69, 104011
\bibitem{Noh-Hwang-Bucher-2019}
         Noh H., Hwang J., Bucher M., 2019, ApJ, 877, 124 (NHB)
\bibitem{Poisson-Will-2014}
         Poisson E, Will C. M., 2014, Gravity: Newtonian, Post-Newtonian, Relativistic (Cambridge Univ. Press)
\bibitem{Racine-Flanagan-2005}
         Racine \'E., Flanagan \'E. \'E., 2005, Phys. Rev. D, 71, 044010
\bibitem{Shibata-Asada-1995}
         Shibata M., Asada H., 1995, Prog. Theor. Phys., 94, 11,
\bibitem{Shu-1992}
         Shu F. H., 1992, The Physics of Astrophysics vol II Gas Dynamics (University Science Books)
\bibitem{Weinberg-1972}
         Weinberg S., 1972, Gravitation and Cosmology (John Wiley \& Sons)
\end{thebibliography}
\end{document}